\documentclass[11pt,a4paper]{article}
\usepackage{amsmath,latexsym,fullpage,amssymb,color}
\usepackage{epic,eepic,ifthen,graphics,epsfig}
\usepackage[english]{babel}
\usepackage{times}
\usepackage{float}
\usepackage{xspace}
\usepackage[T1]{fontenc}
\newlength {\squarewidth}

\newtheorem{theorem}{Theorem}
\newtheorem{lemma}{Lemma}

\newcommand{\toto}{xxx}
\newenvironment{proofT}{\noindent{\bf Proof }}
{\hspace*{\fill}$\Box_{Theorem~\ref{\toto}}$\par\vspace{3mm}}
\newenvironment{proofL}{\noindent{\bf Proof }}
{\hspace*{\fill}$\Box_{Lemma~\ref{\toto}}$\par\vspace{3mm}}

\newfloat{algorithm}{thp}{lop}
\floatname{algorithm}{Algorithm}
\newfloat{construction}{thp}{lop}
\floatname{construction}{Construction}

\newenvironment{lemma-repeat}[1]{\begin{trivlist}
\item[\hspace{\labelsep}{\bf\noindent Lemma~\ref{#1} }]}
{\end{trivlist}}

\newenvironment{theorem-repeat}[1]{\begin{trivlist}
\item[\hspace{\labelsep}{\bf\noindent Theorem~\ref{#1} }]}
{\end{trivlist}}

\newcounter{linecounter}
\newcommand{\linenumbering}{\ifthenelse{\value{linecounter}<10}
{(\arabic{linecounter})}{(\arabic{linecounter})}}
\renewcommand{\line}[1]{\refstepcounter{linecounter}\label{#1}\linenumbering}
\newcommand{\resetline}[1]{\setcounter{linecounter}{0}#1}
\renewcommand{\thelinecounter}{\ifnum \value{linecounter} > 
9 \else \fi\arabic{linecounter}}
\newfloat{algorithm}{thp}{lop}
\floatname{algorithm}{Algorithm}
\newfloat{construction}{thp}{lop}
\floatname{construction}{Construction}
\newcommand{\Xomit}[1]{}
\newcommand{\SM}{\mathit{SM}}
\newcommand{\CT}{\mathit{CT}}

\newcommand{\acquire}{{\sf{acquire}}}
\newcommand{\release}{{\sf{release}}}

\newcommand{\op}{{\sf{op}}}

\newcommand{\mmax}{{\sf{max}}}
\newcommand{\ggcd}{{\sf{gcd}}}
\newcommand{\llog}{{\sf{log}}}

\newcommand{\scan}{{\sf{scan}}}

\newcommand{\map}{{\sf{map}}}

\newcommand{\DESA}{{\sc{desa}}}
\newcommand{\BIT}{{\sc{BIT}}}

\newcommand{\RM}{{\sc{RM}}}

\newcommand{\MUTEX}{{\sc{mutex}}}
\newcommand{\APPL}{{\sc{appl}}}
\newcommand{\desanonymize}{{\sf{desanonymize}}}
\newcommand{\ok}{{\tt{ok}}}

\newcommand{\ttrue}{{\tt{true}}}
\newcommand{\ffalse}{{\tt{false}}}

\begin{document}

\title{\bf Mutex-based Desanonymization of\\
            an  Anonymous  Read/Write Memory}

\author{Emmanuel Godard$^{\dag}$, Damien Imbs$^{\dag}$, 
        Michel Raynal$^{\star,\ddag}$, Gadi Taubenfeld${^\circ}$  \\~\\
$^{\dag}$Aix-Marseille Universit\'e, CNRS, Universit\'e de Toulon, LIS, Marseille, France  \\
$^{\star}$Univ Rennes IRISA, France \\
$^{\ddag}$Department of Computing, Polytechnic University, Hong Kong\\
${^\circ}$The Interdisciplinary Center, Herzliya 46150, Israel
}

\date{}

\maketitle

\begin{abstract}
Anonymous shared memory is a memory in which processes use different
names for the same shared read/write register.  As an example, a shared
register named $A$ by a process $p$ and a shared register named $B$ by
another process $q$ can correspond to the very same register $X$, and
similarly for the names $B$ at $p$ and $A$ at $q$ which can  correspond to
the same register $Y\neq X$. Hence, there is a permanent disagreement on the
register names among the processes.  This new notion of anonymity was
recently introduced by G.\ Taubenfeld (PODC 2017), who presented several memory-anonymous algorithms and impossibility results.

This paper introduces a new problem (new to our knowledge), that
consists in ``desanonymizing'' an anonymous shared memory.  To this
end, it presents an algorithm that, starting with a shared memory made
up of $m$ anonymous read/write atomic registers (i.e., there is no a priori
agreement on their names), allows each process to compute a local
addressing mapping, such that all the processes agree on the names of
each register.  The proposed construction is based on an
underlying deadlock-free mutex algorithm for $n\geq 2$ processes
(recently proposed in a paper co-authored  by some of the authors
of this paper), and consequently inherits its
necessary and sufficient condition  on
the size $m$ of the anonymous memory, namely $m$ must belongs to the set
$M(n)=\{m:~\mbox{ such that }
\forall~ \ell: 1<\ell \leq n:~ \ggcd(\ell,m)=1\}\setminus \{1\}$.
This algorithm, which is also symmetric in the sense process
identities can only be compared by equality, requires the
participation of all the processes; hence it can be part of the  system
initialization.
Last but not least, the proposed
algorithm has a first-class noteworthy property, namely, its simplicity.

~\\~\\{\bf Keywords}:
Anonymity, Anonymous shared memory, Asynchronous system,
Atomic read/write register, Concurrent algorithm,  Deadlock-freedom,
Local memory, Mapping function, Mutual exclusion, Simplicity, Synchronization.
\end{abstract}

\section{Introduction}
\vspace{-0.1cm}
{\bf Read/write registers.}
{\it Read/write registers} are the basic objects of sequential computing.
From a theoretical point of view they constitute the cells of a
Turing machine tape, and from a programming point of view, they are the
memory locations on top of which are built high-level objects such as
stacks, queues, and trees (to cite a few of the most common).

In a concurrent programming context, a read/write register
can be shared (accessed)  by several processes to coordinate their actions or
progress to a common goal.
The most popular consistency condition for registers is {\it atomicity},
which states that all its read and write operations appear as if they
have been executed sequentially, this sequence $S$ being such that,
if an operation $\op1$ terminates before operation $\op2$ starts,
$\op1$ appears  before  $\op2$ in $S$, and a read operation returns
the value written by the closest preceding write in $S$~\cite{L86}.

A register is said to be single-reader (SR) or multi-reader (MR)
according to the number of processes that are allowed to read it.
Similarly, a register can be single-writer (SW) or multi-writer (MW).
A lot of algorithms have been proposed (e.g., see the
textbooks~\cite{R13,T06}), which build MWMR registers from SWSR or SWMR
registers in the presence of asynchrony and process crashes.  In the
other direction, an adaptive construction of  SWMR registers from
MWMR registers is described in~\cite{DFGL13}.\\

\vspace{-0.1cm}
\noindent
{\bf Anonymous memory.}
While the notion of {\it process anonymity} has been studied for a
long time from an algorithmic and computability point of view, both in
message-passing systems (e.g.,~\cite{A80,BR13,YK96}) and shared
memory systems (e.g.,~\cite{AGM02,BRS18,GR07}), the notion of
{\it  memory anonymity} has been  introduced only very recently
in~\cite{T17}. (See~\cite{RC18} for an introductory survey
on  process and memory anonymity).

Let us consider a shared memory $\SM$ made up of $m$ atomic read/write
registers.  Such a memory can be seen as an array with $m$ entries,
namely $\SM[1..m]$.  In a non-anonymous memory system, for each index
$x$, the name $\SM[x]$ denotes the same register whatever the process that
invokes the address $\SM[x]$. As stated in ~\cite{T17},
in the classical system model, there is an a priori agreement on
the names of the shared registers. This a priori agreement facilitates
the implementation of the coordination rules the processes
have to follow to progress without violating the safety (consistency)
properties associated with the application they solve~\cite{R13,T06}.

This a priori agreement does no longer exist in a memory-anonymous
system.  In such a system the very same address identifier $\SM[x]$
invoked by a process $p_i$ and invoked by a different process $p_j$
does not necessarily refer to the same atomic read/write register.
More precisely, a memory-anonymous system is such that:
\begin{itemize}
\item for each process $p_i$
  an adversary defined, over the set $\{1,2,\cdots,m\}$,
  a permutation $f_i()$ such that when $p_i$ uses the address
$\SM[x]$, it actually accesses $\SM[f_i(x)]$, and
\item no process knows the permutations.
\end{itemize}
Let us notice that the read/write registers of a memory-anonymous
system are necessarily MWMR. \\

\vspace{-0.1cm}
\noindent
{\bf Results on anonymous memory.}
In \cite{T17},
mutual exclusion, consensus, and renaming, problems are addressed, and
memory-anonymous algorithms and impossibility results are presented.  Concerning
deadlock-free mutual exclusion in failure-free asynchronous read/write
systems, he presented:
\begin{itemize}
\item
A symmetric deadlock-free algorithm for two processes
(``symmetric'' means process identifiers are not ordered and
can only be compared for equality, see Section~\ref{sec:def-symmetric}).
\item
A theorem stating there is no  deadlock-free algorithm
if the number of processes $n$ is not known.
\item
  A condition on the size $m$ of the anonymous memory
  which is necessary for any  symmetric deadlock-free algorithm.
  More precisely, given an $n$-process system where $n\geq 2$,
  there no deadlock-free mutual exclusion algorithm if the size  $m$
  does not belong to the set  $M(n)=
  \{~m~ \mbox{ such that }\forall~ \ell:~ 1 < \ell \leq n$: $\ggcd(\ell,m)=1\}
  \setminus \{1\}$.
\end{itemize}

  Let us observe that  the previous  condition implies that it is not
  possible to design a symmetric deadlock-free mutex algorithm when
  the size of the anonymous memory $m$ is an even integer greater than $2$.
  As such algorithms can be designed whatever $m$ in a non-anonymous memory,
  it follows that, when the size of the  memory $m$ is an even integer
  greater than $2$, non-anonymous  read/write registers are computationally
  stronger than anonymous  registers.

In the conclusion of \cite{T17}, a few
open problems are presented, one of them being ``the existence of a symmetric starvation-free mutual exclusion algorithms for two processes'',
another one being ``the existence of a symmetric deadlock-free mutual
exclusion algorithm for more than two processes''.  This second
problem was recently solved in~\cite{ARW18} where an algorithm is
presented, which assumes $m\in M(n)$.  It follows that the very
existence of this algorithm shows that the condition $m\in M(n)$ is
also a sufficient condition for symmetric deadlock-free mutual
exclusion in read/write anonymous memory systems.\\

\vspace{-0.1cm}
\noindent
{\bf Content of the paper.}
As shown in~\cite{ARW18,T17}, the design of memory-anonymous
algorithms is not a trivial task.
We started this work with an attempt to design a
starvation-free memory-anonymous mutual exclusion algorithm.  This
drove us to the observation that the fact ``there is currently a
competition among processes'' must be memorized in one way or another
to prevent a process from always defeating other processes, and
thereby ensure starvation-freedom.

Finally, considering an $n$-process system, after many attempts, this
work ended with a relatively simple symmetric {\it desanonymization}
algorithm,
namely, an algorithm that transforms an anonymous read/write memory
into a non-anonymous read/write memory.
This algorithm requires the participation of all the processes,
and assumes that processes do not fail.
  Once  memory desanonymization is obtained (e.g., at system initialization),
  it becomes possible to
use algorithms based on a non-anonymous memory on top of an anonymous
memory.

The proposed construction relies on an underlying memory-anonymous
symmetric deadlock-free mutual exclusion algorithm (the one introduced
in~\cite{ARW18}). Hence, it inherits its requirement on $m$, namely,
$m\in M(n)$.
It follows that, when $m$ satisfies this condition, $m$ anonymous
registers and $m$ non-anonymous registers have the same computability
power from an anonymous/non-anonymous mutual exclusion point of
view. Let us also notice that, if a non-anonymous memory algorithm
executed on top of the proposed construction requires $m'$ registers
where $m'$
does not belong to the set $M(n)$ defined above, it is
sufficient to select the first integer greater than $m'$ belonging to
$M(n)$  as the value of $m$, and, at the non-anonymous
memory upper layer, $(m-m')$ registers are ignored.
Let us notice that  the  proposed construction is {\it universal} in the
sense any concurrent  non-anonymous memory algorithm can be
executed on top of it.\\

\vspace{-0.1cm}
\noindent
{\bf On the difficulty of the problem.}
In a non-anonymous memory system, the
read/write registers used by an algorithm are accessed only by
this algorithm.  Its identifiers are unambiguously shared by
all processes, and no other algorithm is allowed to concurrently use
these registers.
Differently, in an anonymous  memory system,  a process must
(in one way or another) write ``enough''  registers to transmit
information to other processes. This is a direct consequence of the
fact that  there is no a priori agreement on the identities of the shared
atomic read/write registers and the fact that -- due to its very nature --
no anonymous register can be a single-writer register.

Hence, the difficulty in the construction of a memory desanonymization
algorithm comes from the fact that, due to memory anonymity, it
concurrently uses the same registers as the ones used by
the underlying mutex algorithm it uses as a subroutine.
As we will see, to circumvent this issue,
the proposed memory desanonymization algorithm will
use (in a very simple way) the local
memory of each process to store the value of an increasing counter,
which simulates a  shared non-anonymous register on which the processes
agree and can consequently use to coordinate their local progress.

The desanonymization problem addressed in this paper may seem of
theoretical interest only (as many other problems appeared first).
As long as its practical interest is concerned, we do not have to forget
that, as nicely expressed  by the physicist Niels Bohr ``prediction is
very difficult, especially when it about the future!''.
Nevertheless, the results presented in this paper shows
that, from a computability point of view,
there are cases where --in a failure-free context-- anonymous read/write
registers are as strong as non-anonymous  registers.

Let us also notice that a similar problem (but much simpler, even
trivial) appears in message-passing systems, where any two nodes
(processes) are connected by a communication channel, locally known as
internal ports by each process, $port_i[x]$ being the local name of
the channel connecting process $p_i$ to some process $p_j$.  In this
context, it is possible that for any two processes $p_i$ and $p_k$,
the local names $port_i[x]$ and $port_k[x]$ denote channels connecting
them to two different processes, while $port_i[x]$ and $port_k[y]$,
$x\neq y$, connect them to the same process. Differently from process
identities, values stored in ports are purely local and have no global
meaning.  Moreover, it is straightforward for a process to learn the
name of the process it is connected to when it uses a given local
port.\\

\vspace{-0.1cm}
\noindent
{\bf Simplicity is a first class property.}
The simplicity of the proposed algorithm does not mean it was simple
to obtain. This was not a trivial task as simplicity is rarely
obtained for free.  As said by A.J. Perlis (the first Turing Award recipient)
``Simplicity does not precede complexity, but follows it''~\cite{P82}.
Let us also remember the following sentence written by the
mathematician/philosopher Blaise Pascal at the end of a letter to a
friend: ``I apologize for having written such a long letter, I had not
enough time to write a shorter one''. The implication ``simple
$\Rightarrow$ easy'' is rarely true for non-trivial
problems~\cite{AZ10}.  Simplicity requires effort, but is very
rewarding. It is a first class scientific property which participates
in the beauty of science~\cite{D80}.\\

\vspace{-0.1cm}
\noindent
{\bf Roadmap.}
The paper is composed of~\ref{sec:conclusion} sections.
Section~\ref{sec:model} introduces the computing model, the
notion of a symmetric algorithm, and mutual exclusion.
Section~\ref{sec:def-desa} defines the  desanonymization problem.
A  first  desanonymization  algorithm is presented in
Section~\ref{sec:algorithm} and proved in Section~\ref{sec:proof}.
This algorithm  requires each register of the desanonymized memory
to forever contain $1+\llog_2 m$ bits of control information.
Then, the previous algorithm is enriched in Section~\ref{sec:one-bit-desa}
to obtain  an algorithm which associates a single bit of permanent
control information with each register of the desanonymized memory.
Section~\ref{sec:conclusion} concludes the paper.

\section{System Model, Symmetric Algorithm, and Mutex  Algorithm}
\label{sec:model}
\subsection{Process and Communication Model}
\label{sec:memory-anonymous}
\noindent
{\bf Processes.}
The system is composed of a finite set of $n\geq 2$ asynchronous
processes denoted $p_1$, .., $p_n$. The subscript $i$ in $p_i$ is only
a notational convenience, which is not known by the processes.
{\it Asynchronous} means that each process proceeds to its own speed,
which can vary with time and remains always unknown to the other
processes. Each process $p_i$ knows its identity $id_i$ and the total
number of processes $n$. No two processes have the same identity, and
this is known  by all processes. \\

\noindent
{\bf Anonymous shared memory.}
The shared memory is made up of $m$ atomic anonymous read/write
registers denoted $\SM[1...m]$. Hence, {\it all} registers are
anonymous.  As indicated in the Introduction, when $p_i$ uses the
address $\SM[x]$, it actually uses $\SM[f_i(x)]$, where $f_i()$ is a
permutation defined by an external adversary.  We will use the
notation $\SM_i[x]$ to denote $\SM[f_i(x)]$, to stress the fact that
no process knows the permutations.

It is assumed that all the registers are initialized to the same value.
Otherwise, thanks  to their different initial values, it would be possible
to distinguish different registers, which consequently will no longer
be fully anonymous. \\

\noindent
{\bf To summarize: which adversaries?}
The adversaries considered in the paper are consequently asynchrony
and memory anonymity. There are no process failures (this assumption is
motivated by the fact that the proposed construction is based on a
mutual exclusion algorithm, and mutual exclusion algorithms are
impossible to build from read/write registers in the presence of
process failures).
Furthermore, unlike the mutual exclusion model where a process may never leave its remainder region, we assume that all the processes must participate in the algorithm.

\subsection{Symmetric Algorithm}
\label{sec:def-symmetric}
The notion of a {\it symmetric algorithm} dates back to the
eighties~\cite{GG90,JS85}. Here, as in~\cite{T17}, a {\it symmetric
  algorithm} is an ``algorithm in which the processes are executing
exactly the same code and the only way for distinguishing processes is
by comparing identifiers. Identifiers can be written, read, and
compared, but there is no way of looking inside an identifier. Thus it
is not possible to know  whether an identifier is odd or even''.

Moreover, symmetry can be restricted by considering that the only
comparison that can be  applied to identifiers is equality. In this case,
there is no order structuring the identifier name space.
In the following, we consider the more restricting definition, namely,
``symmetric'' means `` symmetric with comparison limited to equality''.

Let us notice that, as all the processes have the same code and all the
registers are initialized to the same value, process identities become
a key element when one has to design an algorithm in such a constrained context.

\subsection{One-Shot Mutual Exclusion}
\label{sec:mutex-one-shot}
\noindent
{\bf One-Shot Mutual Exclusion.}
Mutual exclusion is the oldest (and one of the most important) synchronization
problem.  Formalized by E.W. Dijkstra in the mid-sixties~\cite{D65},
it consists in building what is called a lock (or mutex) object,
defined by two operations, denoted $\acquire()$ and $\release()$.
(Recent textbooks including mutual exclusion and variants of it
are~\cite{R13,T06}.)

The invocation of these operations by a process $p_i$
always follows the following pattern: ``$\acquire()$; {\it critical section};
$\release()$'', where ``critical section'' is any  sequence of code.
Moreover, ``one-shot'' means that a process invokes at most once
the operations $\acquire()$ and $\release()$.
The mutex object satisfying the deadlock-freedom progress condition
is defined by the following two properties.
\begin{itemize}
\item
Mutual exclusion.
No two processes are simultaneously in their critical section.
\item
Deadlock-freedom progress condition.  If there is a process
$p_i$ that has a pending operation $\acquire()$, there is a process $p_j$
(maybe $p_j\neq p_i$) that eventually executes its critical section.
\end{itemize}
As already mentioned, a
 memory-anonymous symmetric deadlock-free mutual exclusion algorithm
is presented in~\cite{ARW18}.  This algorithm assumes that size $m$ of
the anonymous memory
belongs to the set $M(n)= \{~m~\mbox{ such that }
\forall \ell:~ 1 < \ell \leq n$: $\ggcd(\ell,m)=1\}\setminus \{1\}$.
Hence, the
mutex-based read/write memory desanonymization algorithm presented in
Section~\ref{sec:algorithm} is optimal with respect to the values of
$m$ for which deadlock-free mutual exclusion can be built despite
memory anonymity. \\

\noindent
{\bf One-Shot Mutual Exclusion vs Election.}
One-shot mutual exclusion and election are close but different
problems.  The operations $\acquire()$ and $\release()$ (which allows
a process to ``bracket'' a sequence of code) allows us to exploit the
order in which processes enter their critical section.
An election object provides the processes with a single operation
${\sf elect}()$, which returns true to exactly one process
and false to all other processes.

Hence, an election instance does not allow to totally order the whole set of
processes.  Differently, a one-shot mutual exclusion instance allows
to order the processes, namely, in the order in which they
enter the critical section.  It follows that one-shot mutual exclusion
is strictly stronger than election in the sense that an instance of
one-shot mutual exclusion allows solving election, while the opposite
is not true.

\section{The Desanonymization Problem}
\label{sec:def-desa}

\noindent
{\bf Definition.}
Given an $n$-process  asynchronous system, in which the processes
communicate via set of $m$ anonymous read/write registers $\SM[1..m]$,
the aim is for each process $p_i$  to compute an addressing function
$\map_i()$, which is a permutation over the set of the memory indexes
$\{1,\cdots,m\}$, such  that  the two following properties are satisfied.
It is assumed that all processes participate in the  desanonymization.
\begin{itemize}
\item Safety.
  For any  $y \in \{1,\cdots,m\}$ and any process $p_i$, we have
  $\SM_i[\map_i(y)]= \SM[y]$.
\item Liveness.  There is a finite time after which all the processes
  have computed their addressing function $\map_i()$.
\end{itemize}
The safety property states that, once a process $p_i$ has computed
$\map_i()$, its local anonymous memory address $\SM_i[x]$, where
$x=\map_i(y)$, denotes the  shared register $\SM[y]$.\\

\begin{figure}[htb]
\centering{
  \scalebox{0.25}{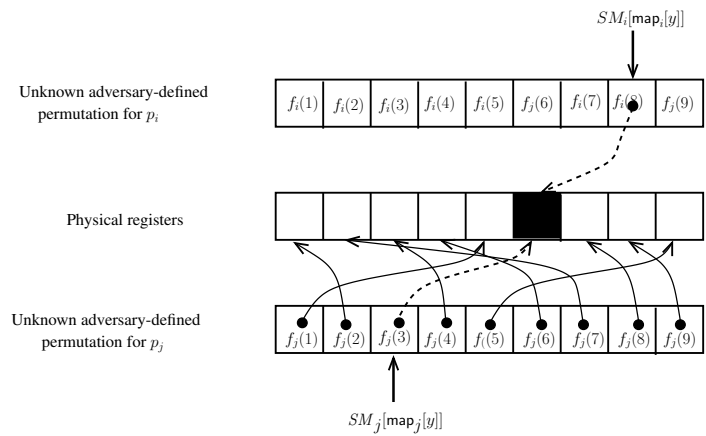}
\caption{Accessing the memory after desanomyzation}
\label{fig:mapping-example}
}
\end{figure}

\noindent
{\bf Accessing the desanonymized memory.}  Once  desanonymized, the way 
the memory is accessed by the processes is illustrated in
Fig.~\ref{fig:mapping-example}. For any index $y$, $1\leq y\leq m$,
the processes access the same register as follow: $\SM_i[\map_i[y]]$
used by $p_i$ and $\SM_j[\map_j[y]]$ used by $p_j$ denote the same register.
\vspace{-0.2cm}
\vspace{-0.2cm}

\section{A Symmetric  Desanonymization Algorithm}
\label{sec:algorithm}

\subsection{Memory Desanonymization in an  $n$-Process Read/Write System}

\noindent
{\bf Underlying principle.}
The principle that underlies the design of the read/write
memory desanonymization
algorithm (Algorithm~\ref{algo:n-process-desanonymization}) is based
on an {\it competition/elimination} process, at the end of which a
single winner process imposes its adversary-defined index permutation
to all the processes, which becomes the shared names of the anonymous
read/write registers, on which all processes agree.

The competition/elimination process uses an underlying mutual
exclusion algorithm.  Each process invokes $\acquire()$ and is
eliminated when it leaves the critical section. The last process to
enter the critical section is the winner.\\

\noindent
{\bf Challenges.}
In order to detect which process is the last, the processes needs to
collaborate to increase a counter whose value will reach $n$ when the
last process will enter the critical section.  We stress that because
the memory is anonymous there is no straightforward way to leverage a
critical section. Since there is no agreement on the resources (here
the anonymous registers themselves), being in critical section does
not grant any restricted access to the memory.  In the following,
properties of the underlying algorithm are described, which are used
to build the required shared resource, namely a shared counter.\\

\noindent
{\bf Properties of the underlying mutex  algorithm that are used.}
In addition to the fact it solves mutual exclusion,
the underlying mutex algorithm has behavioral properties that are
implicitly used in the design of the
desanonymization algorithm and explicitly used in its proof.

\begin{itemize}
\item Property Mutex-1.
  A process writes only its identity or $\bot$ in an anonymous register.
\item Property Mutex-2.
When a process invokes  $\acquire()$, it reads all anonymous registers.
\item Property Mutex-3.
  When a process is allowed to enter the critical section,
      all registers contain its identity.
\item Property Mutex-4.
  After a process is allowed to enter the critical section
  and before it invokes $\release()$,
  any other competing  process can issue at most one write operation.
  It follows that, when a process $p_i$ is inside the critical section,
  and $x$ processes are inside their invocations of $\acquire()$,
  at least $(m-x)$ anonymous registers  contain its identity $id_i$.
  Moreover, when a process release the critical section (operation
  $\release()$), it writes $\bot$, in all the registers which contain
  its identity. Hence, at least $(m-x)$ such registers are reset to
  their initial value $\bot$.
\end{itemize}

\noindent
{\bf Enriching the underlying mutex algorithm to share a counter.}
As can be seen from the previous properties, even when a process is
alone in the critical section, it could happen that some of its
writes are overwritten by another process. Property Mutex-4 states that a
process, which is not in the critical section, may erase what was written by
the process in critical section only once. That is no more than  $(n-1)$
registers can  be erased.
As $m-(n-1)>0$,   by copying the value in {\it all} the anonymous
registers, the process currently in
the critical section ensures that at least one copy will not be
overwritten. From property Mutex-2, the next process to enter the
critical section will learn the correct value of the counter.

Sharing the counter in such a way is more easily done by integrating
these operations within each read and write operation on the anonymous
registers,  issued by the underlying mutual exclusion
algorithm. These basic operations are consequently enriched as described in
Algorithm~\ref{algo:rw-mutex}. These modifications are safe for the
mutual exclusion algorithm since they do not interfere with operations
and variables of this algorithm.

Let us remark that a similar technique, based on appropriate broadcast
abstraction and quorums,  is used in message-passing systems
to update the local copies of a shared register~\cite{R18}.
Here the read and write operations issued by the underlying mutex algorithm
are enriched to play the role of a broadcast abstraction.\\

\noindent
{\bf Local variables.}
Each process s $p_i$ manages three local  variables.
\begin{itemize}
\item
$ct_i$ is a local counter initialized to $0$, which will increase inside
  the integer interval $[0..n]$.  The set of the $n$ local variables
  $ct_i$ implement a shared counter $\CT$ which increases by step $1$
  from its initial value $0$ to $n$ (line~\ref{DESA-n-proc-02}).
  (Actually, the set of the final values of the $n$ local variables $ct_i$
  will be the set $\{1, 2, \ldots,n\}$.)
\item
  $sm_i[1..m ]$ is used to store a local copy of the anonymous memory
  $SM_i[1..m]$. A process $p_i$ reads the anonymous memory by invoking
  $\SM_i.\scan()$, which is an asynchronous (non-atomic) reading of
  all the anonymous registers.
\item $last1_i$ is a Boolean, initialized to false, which will be set to
  true only by the last process that will access the critical section.
\end{itemize}

\noindent
{\bf Each register contains a tag and  a value.}
In order not to confuse the values written in anonymous registers by
processes executing statements of
Algorithm~\ref{algo:n-process-desanonymization} (not including the
operations $\acquire()$ and $\release()$), and the values written by
other processes executing the underlying mutex algorithm, all the
values written in the anonymous memory are prefixed by a tag.  More
explicitly, the tag \MUTEX\ is used by the mutex algorithm, while the
tag \DESA\ is used by the desanonymization algorithm.

Each anonymous read/write  register is initialized to
\MUTEX$\langle 0,\bot\rangle$.  The first value ($0$) is the
initial value of the global counter $\CT$, while the second value
($\bot$) is the initial value used by the mutex algorithm.

\begin{algorithm}[htb]
\centering{
\fbox{
\begin{minipage}[t]{150mm}
\footnotesize
\renewcommand{\baselinestretch}{2.5}
\resetline
\begin{tabbing}
aaa\=a\=aaa\=aaa\=aaaaa\=aaaaa\=aaaaaaaaaaaaaa\=aaaaa\=\kill

{\bf operation} $\SM_i.\scan()$
{\bf returns} $([SM_i[1], \cdots,SM_i[m]])$.\\~\\

{\bf operation}  $\desanonymize(id_i)$ {\bf is}
    $~~~~$ \% code for process $p_i$ \\

\line{DESA-n-proc-01} \> $\acquire(id_i)$; \\

\line{DESA-n-proc-02} \> \> $ct_i \leftarrow ct_i+1$;\\
\>  \> \%  $ct_i$ is the local representation of the global counter $CT$.
     It is updated at each process   \\
\>  \> \%  by the read and write operations of the underlying mutex algorithm
(see Algorithm~\ref{algo:rw-mutex})  \\

\line{DESA-n-proc-03}
\>  \>  $last1_i \leftarrow (ct_i=n)$;\\

\line{DESA-n-proc-04}
 \> $\release(id_i)$;  \% realizes an implicit broadcast of $ct_i$  \% \\

\line{DESA-n-proc-05} \> {\bf if} $(last1_i)$ \\

\line{DESA-n-proc-06}
\>\> {\bf th}\={\bf en} {\bf for each} $x\in\{1, \cdots,m\}$ {\bf do}
 $\SM_i[x] \leftarrow$ \DESA$(x)$ {\bf end for}\\

\>\> \>  \%  the permutation for $p_i$ is:
  $\forall~ y\in\{1,\cdots,m\}$: $\map_i(y)=y$  \%\\

\line{DESA-n-proc-07}
\>\> {\bf else} \= {\bf repeat}  \= $sm_i\leftarrow  \SM_i.\scan()$
  {\bf until} ($\forall ~x:~  sm_i[x]$ is tagged  \DESA)
        {\bf end repeat};\\

\line{DESA-n-proc-08}
\>\>\> {\bf for each} $x\in\{1, \cdots,m\}$ {\bf do}
                  $\map_i(y)\leftarrow x$ where  $sm_i[x]$=\DESA$(y)$
       {\bf end for}\\

\>\> \>   \%  the perm. for $p_i$ is:
        $\forall~y \in\{1,\cdots,m\}$:
$\map_i(y)= x$,  where $sm_i[x]=$ \DESA$(y)$\\

\line{DESA-n-proc-09}  \> {\bf end if}.

\end{tabbing}
\normalsize
\end{minipage}
}
\caption{Memory desanonymization in an $n$-process read/write system}
\label{algo:n-process-desanonymization}
}
\end{algorithm}

\noindent
{\bf Behavior of a process $p_i$: first invoke the mutex algorithm.}
All the processes invoke the operation $\desanonymize(id_i)$.  When a
process $p_i$ invokes it, it first acquires the critical section
(line~\ref{DESA-n-proc-01}).  The code inside the critical section is
a simple increase of the shared counter $\CT$ globally implemented by
the local variables $ct_i$ (line~\ref{DESA-n-proc-02}).  Hence, if
$p_i$ is the $\ell^{\mbox{th}}$ process to access the critical section, $ct_i$
is updated from $\ell-1$ to $\ell$, and $p_i$ will inform the other
processes of this increase when it will invoke $\release()$
(line~\ref{DESA-n-proc-04}).  Let us notice that, at
line~\ref{DESA-n-proc-03}, $p_i$ sets to true its local Boolean variable
$last1_i$ only if it is the last process to execute the critical section.
Then, the behavior of $p_i$ depends on the fact it is or not the
last process to enter the critical section (see below). \\

\noindent
{\bf Behavior of a process $p_i$: the read and write
operations used by the  mutex algorithm.}
As already indicated, to ensure a correct dissemination of the last increase of
$\CT$ (update of the local variable $ct_j$ at a process $p_j$),
the read and write operations that allow the mutex algorithm
to  access the  anonymous registers are modified as described in
Algorithm~\ref{algo:rw-mutex}.

\begin{algorithm}[h!]
\centering{
\fbox{
\begin{minipage}[t]{150mm}
\footnotesize
\renewcommand{\baselinestretch}{2.5}
\resetline
\begin{tabbing}
aaaaa\=aa\=aaa\=aaa\=aaaaa\=aaaaa\=aaaaaaaaaaaaaa\=aaaaa\=\kill

{\bf operation} 
               read of $\SM_i[x]$ executed by the mutex algorithm {\bf is} \\

\line{MUTEX-RW-01} \> $\langle ct, val\rangle \leftarrow SM_i[x]$;\\

\line{MUTEX-RW-02}\> $ct_i \leftarrow \mmax(ct_i,ct)$;\\

\line{MUTEX-RW-03} \>${\sf return} (val)$.\\~\\

{\bf operation} 
 write of $v$ in  $\SM_i[x]$ executed by the mutex algorithm {\bf is} \\

\line{MUTEX-RW-04} \> $SM_i[x] \leftarrow$ \MUTEX$\langle ct_i, v\rangle$;\\

\line{MUTEX-RW-05} \> ${\sf return} (\ok)$.

\end{tabbing}
\normalsize
\end{minipage}
}
\caption{Modified read and write operations
  (code for $p_i$)}
\label{algo:rw-mutex}
}
\end{algorithm}

As the operation $\release()$ of the mutex algorithm writes $\bot$
(i.e., the \MUTEX$\langle\CT,\bot\rangle$) in at least $(m-(n-1))$
anonymous registers (Property Mutex-4), it follows that if a process
$p_i$ accesses later the critical section, it updated its local
counter $ct_i$ when it executed $\acquire()$, which reads all
anonymous registers (Property Mutex-1).\\

\noindent
{\bf Behavior of a process $p_i$: the winner imposes its
           addressing permutation to all.}
The desanonymization is done at lines~\ref{DESA-n-proc-05}-\ref{DESA-n-proc-09}.
The $(n-1)$ processes that won the first  $(n-1)$ critical sections
execute line~\ref{DESA-n-proc-07}, in which they
loop until they see all the registers tagged \DESA.

Let $p_\ell$ be the last process that entered the critical section
(hence, $ct_\ell=n$ and $last_\ell$ is the only Boolean equal to
true).  This process imposes its adversary-defined addressing
permutation as the common addressing, which realizes a non-anonymous
memory. To this end, for any $x\in\{1,\cdots,m\}$, $p_\ell$ writes
\DESA$(x)$ in $SM_\ell[x]$ (line~\ref{DESA-n-proc-06}).  Hence, for
any $x$ we have $\map_\ell(x)=x$.

Let $p_i$ be any other process that is looping at
line~\ref{DESA-n-proc-07} until it sees all the registers tagged \DESA.
When this occurs, it computes $\map_i()$, which is such that for any
$x\in\{1,\cdots,m\}$, if $sm_i[x]=$\DESA$(y)$ then $\map_i(x)=y$
(line~\ref{DESA-n-proc-07}).

\subsection{Using the Desanonymized Memory}
\label{sec:using-desa-memory}

It follows from the desanonymization algorithm that when a process has
written the tag \DESA\ in all registers, thanks to their local mapping
function $\map_i()$, all the processes share the same indexes for the
same registers.

When this occurs, process $p_k$ could start executing its local
algorithm defined by the upper layer application, but if it writes an
application-related value in some of these registers, this value can
overwrite a  value \DESA$()$ stored in a register not yet read by
other processes.
To prevent this problem from occurring, all the values written by a
process at the application level are prefixed by the tag \APPL, and
include a field containing the common index $y$ associated with this
register. In this way, any process $p_i$ will be able to compute its
local mapping function $\map_i()$, and can start its upper layer
application part, as soon as it has computed $\map_i()$.

Let us notice that one bit is needed to distinguish the tag  \DESA\
and the tag \APPL. Hence, each of a value  \DESA$(x)$  and a value
\APPL$(x,-)$ requires  $(1+\llog_2~m)$ control  bits.

\section{Proof of the Algorithm}
\label{sec:proof}

\vspace{-0.1cm}
\begin{lemma}
  \label{lemma-counter}
  Each process exits $\acquire()$ and, denoting ${i_k}$ the index
  of the  $k^{\mbox{th}}$  process that  enters the critical section,
  when $p_{i_k}$ invokes $\release()$, it writes the value
  \MUTEX$\langle k,-\rangle$ in at least $(m-(n-1))$ anonymous registers.
\end{lemma}

\vspace{-0.1cm}
\begin{proofL}
 Let us first observe that, as (i) the underlying mutex algorithm is
 independent of the values of the local  variables $ct_i$, (ii) is
 deadlock-free, and (iii) each process invokes $\acquire()$ only once,
 it is actually starvation-free.

  Let $p_{i_1}$ be the first process that enters the critical section.
  As  $ct_{i_1}=0$, it follows that after
  line~\ref{DESA-n-proc-02} we have   $ct_{i_1}=1$.
  Then, when  $p_{i_1}$ invokes $\release()$, it  writes
  \MUTEX$\langle 1,-\rangle$ in at least $(m-(n-1))$ anonymous  registers
  (Property Mutex-4 and
  line~\ref{MUTEX-RW-04} of Algorithm~\ref{algo:rw-mutex}).
  It follows then (i) from Property Mutex-2 and
  lines~\ref{MUTEX-RW-01}-\ref{MUTEX-RW-02} of
  Algorithm~\ref{algo:rw-mutex}),
  and  (ii)  Property Mutex-1,  Property Mutex-3, and
 line~\ref{MUTEX-RW-04} of  Algorithm~\ref{algo:rw-mutex}, that when
  another  process  $p_{i_2}$ enters the critical section,
  $p_{i_2}$ has previously read and  written  all registers,
  from which we conclude from lines~\ref{MUTEX-RW-01}-\ref{MUTEX-RW-05} of
  Algorithm~\ref{algo:rw-mutex} that  $ct_{i_2}=1$.
  It follows that  $p_{i_2}$ increases $ct_{i_2}$ from $1$ to $2$ at
  line~\ref{DESA-n-proc-02} of Algorithm~\ref{algo:n-process-desanonymization}.

  The previous reasoning being repeated $n$ times, we eventually
  have: $ct_{i(x)}=x$ at each process $p_{i(x)}$, $1\leq x\leq n-1$, and
  $ct_{i_n}=n$ at process $p_{i_n}$.  It follows that no process
  blocks forever when it executes the
  lines~\ref{DESA-n-proc-01}-\ref{DESA-n-proc-04}
  of Algorithm~\ref{algo:n-process-desanonymization}.
  \renewcommand{\toto}{lemma-counter}
\end{proofL}

\vspace{-0.1cm}
\begin{lemma}
\label{lemma-safety}
The local mapping function $\map_i()$ computed by each process $p_i$
is a  permutation over the set of register indexes $\{1,\cdots,m\}$.
Moreover,  for any index $y\in\{1,\cdots,m\}$
and any pair of processes $p_i$ and $p_j$,
$\SM_i[\map_i(y)]$ and $\SM_j[\map_j(y)]$ address the very same register.
\end{lemma}
 \vspace{-0.1cm}
\begin{proofL}
  Let us assume that a process $p_i$ executes
  line~\ref{DESA-n-proc-06}.  From Lemma~\ref{lemma-counter} there is
  a single such process $p_i$. Let $p_j$ be any other process that
  executes lines~\ref{DESA-n-proc-07}-\ref{DESA-n-proc-08}.  Due to
  the ``repeat'' loop of line~\ref{DESA-n-proc-07}, $p_j$ executes
  line~\ref{DESA-n-proc-08} only after all registers contain the tag
  \DESA.  Only $p_i$ writes the registers with this tag, and (at
  line~\ref{DESA-n-proc-06}) wrote \DESA$(y)$ inside $\SM_i[y]$, for
  each $y\in\{1,...,m\}$.  Hence, when $p_j$ reads \DESA$(y)$ from
  $\SM_j[x]$, it learns that this register is known by $p_i$ as
  $\SM_i[y]$.  At line~\ref{DESA-n-proc-08}, $p_j$ consequently
  considers $x$ as the value of $\map_j(y)$. It follows that
  $\SM_j[\map_j(y)]$ (i.e., $\SM_j[x]$) and $\SM_i[\map_i(y)]$ (which
  is $\SM_i[y]$) denote the very same read/write register. As this is
  true for any process $p_j\neq p_i$, the lemma follows.
  \renewcommand{\toto}{lemma-safety}
\end{proofL}

 \vspace{-0.1cm}
\begin{lemma}
\label{lemma-termination}
Any process $p_i$ terminates the operation $\desanonymize()$.
\end{lemma}
 \vspace{-0.1cm}
\begin{proofL}
  The proof follows from Lemma~\ref{lemma-counter}, which states that
  all processes enter and leave the critical section. Moreover, as
  $p_{i_n}$ executes line~\ref{DESA-n-proc-06} of
  Algorithm~\ref{algo:n-process-desanonymization}, it follows that no
  other process can block forever at line~\ref{DESA-n-proc-07} of this
  algorithm, which concludes the proof of the lemma.
  \renewcommand{\toto}{lemma-termination}
\end{proofL}

 \vspace{-0.1cm}
\begin{theorem}
\label{theo-main}
Algorithm~{\em\ref{algo:n-process-desanonymization}} is a
symmetric algorithm that
solves the {\em desanonymization} problem in a
system made up of $n$ asynchronous processes communicating by reading
and writing $m$ anonymous read/write atomic registers, where $m$
belongs to the set
$M(n)=\{m\mbox{ such that }\forall \ell:~ 1 < \ell \leq n$: $\ggcd(\ell,m)=1\}
\setminus \{1\}$.
\end{theorem}
 \vspace{-0.1cm}
\begin{proofT}
A simple examination of the code shows that process identities are
compared only by equality, from which follows the ``symmetry''
property.  The rest of the proof follows from Lemma~\ref{lemma-safety}
and Lemma~\ref{lemma-termination}.
\renewcommand{\toto}{theo-main}
\end{proofT}
\vspace{-0.4cm}

\section{Reducing the Size of Control Information}
\label{sec:one-bit-desa}
 \vspace{-0.1cm}
Algorithm~\ref{algo:n-process-desanonymization}
requires that, once desanonymized,  each register must
contain forever  $1+\llog_2m$ bits of control information.
This section shows that this information can be reduced to a single bit.   \\

\noindent
{\bf Revisiting the shared memory.}
Each read/write register $\SM[x]$ is now assumed to be composed of two parts
$\SM[x].\BIT$ and  $\SM[x].\RM$, more precisely, we have
$\SM[x]=\langle\SM[x].\BIT,~\SM[x].\RM\rangle$.
$\SM[x].\BIT$ is for example the leftmost bit of  $\SM[x]$, and
$\SM[x].\RM$ the other bits.
The meaning and the use of $\SM[x].\RM$ are exactly the same as
$\SM[x]$ in Algorithm~\ref{algo:n-process-desanonymization}
and Algorithm~\ref{algo:rw-mutex}.
For each $x$, $\SM[x].\BIT$ is initialized to $0$, while (as in
Algorithm~\ref{algo:n-process-desanonymization}) $\SM[x].\RM$
is initialized to {\sc mutex}$\langle 0,\bot\rangle$.

\begin{algorithm}[htb]
\centering{
\fbox{
\begin{minipage}[t]{150mm}
\footnotesize
\renewcommand{\baselinestretch}{2.5}
\resetline  
\begin{tabbing}
  aaaaa\=aa\=aaa\=aaa\=aaaaa\=aaaaa\=aaaaaaaaaaaaaa\=aaaaa\=\kill 

{\bf operation} $\SM_i.\scan()$
{\bf returns} $([SM_i[1], \cdots,SM_i[m]])$.\\~\\

{\bf operation}  $\desanonymize2(id_i)$ {\bf is} $~~~~$\% code for $p_i$ \\

\> \%  the lines~\ref{BITDESA-01}-\ref{BITDESA-09} are the same as in
Algorithm~\ref{algo:n-process-desanonymization};
  the lines~\ref{BITDESA-10}-\ref{BITDESA-17} are new \\

\line{BITDESA-01} \> $\acquire(id_i)$; \\

\line{BITDESA-02} \> \> $ct_i \leftarrow ct_i+1$;\\

\line{BITDESA-03}
\> \>  $last1_i \leftarrow (ct_i=n)$;\\

\line{BITDESA-04}
\> $\release(id_i)$;  \% realizes an implicit broadcast of $ct_i$  \% \\

\line{BITDESA-05} \> {\bf if} $(last1_i)$ \\

\line{BITDESA-06}
\>\> {\bf then} \= {\bf for each} $x\in\{1, \cdots,m\}$ {\bf do}
 $\SM_i[x] \leftarrow$ \DESA$(x)$ {\bf end for}\\

\>\>\>   \%  the permutation for $p_i$ is:
$\forall~ y\in\{1, \cdots,m\}$: $\map_i(y)=y$  \%\\

\line{BITDESA-07}
\>\> {\bf else} \=
   {\bf repeat}  \= $sm_i\leftarrow  \SM_i.\scan()$
      {\bf until} ($\forall ~x:~  sm_i[x]$ is tagged  \DESA)
         {\bf end repeat};\\

\line{BITDESA-08}
\>\>\> {\bf for each} $x\in\{1, \cdots,m\}$ {\bf do}
                  $\map_i(y)\leftarrow x$ where  $sm_i[x]$=\DESA$(y)$
       {\bf end for}\\

\>\> \>  \%  perm. for $p_i$ is
          $\forall~y \in\{1,\cdots,m\}$:
         $\map_i(y)= x$,  where $sm_i[x]=$ \DESA$(y)$\\

\line{BITDESA-09}  \> {\bf end if};\\

\line{BITDESA-10}  \> $\acquire(id_i)$; \\

\line{BITDESA-11}   \>\> $ct_i \leftarrow ct_i+1$;\\

\line{BITDESA-12}   \>\> $last2_i \leftarrow (ct_i=2n)$;\\

\line{BITDESA-13}
\> $\release(id_i)$;  \% realizes an implicit broadcast of $ct_i$  \% \\

\line{BITDESA-14} \> {\bf if} $(last2_i)$ \\

\line{BITDESA-15}
\>\> {\bf then} \= {\bf for each} $x\in\{1, \cdots,m\}$ {\bf do}
$\BIT_i[x] \leftarrow 1$ {\bf end for}\\

\line{BITDESA-16}
\>\> {\bf else} \> {\bf repeat}  $bit_i\leftarrow  \BIT_i.\scan()$
     {\bf until} ($\exists ~x:~  bit_i[x]=1$)
     {\bf end repeat}\\

\line{BITDESA-17}  \> {\bf end if}.
\end{tabbing}
\normalsize
\end{minipage}
}
\caption{Algorithm with a single bit of control information}
\label{algo:n-process-desa-final-v}
}
\end{algorithm}

To simplify both the writing  and the reading of the
improved algorithm, we  write
\begin{itemize}
\item
`` $\SM_i[x] \leftarrow$ \DESA$(x)$''
  when the first bit of  $\SM_i[x]$ is not modified by the write
(line~\ref{BITDESA-06}),
\item
  ``$\SM_i.\scan()$ when we are interested
  in  the $\SM_i.\RM$'' part of the registers only (line~\ref{BITDESA-07}),
\item
  ``$\BIT_i[x] \leftarrow 1$'' when the
remaining part of
$\SM_i[x]$ is not modified by the write (line~\ref{BITDESA-15}),
\item   ``$\BIT_i.\scan()$'' when we are interested in the bits
 $\SM_i.\BIT$  only  (line~\ref{BITDESA-16}).
\end{itemize}

 \noindent
{\bf Behavior of a process $p_i$.}
Algorithm~\ref{algo:n-process-desa-final-v} is the improved algorithm.
It is Algorithm~\ref{algo:n-process-desanonymization}
(lines~\ref{BITDESA-01}-\ref{BITDESA-09}), followed by a second
global synchronization phase (lines~\ref{BITDESA-10}-\ref{BITDESA-17}),
which is similar to the one at lines~\ref{BITDESA-01}-\ref{BITDESA-09}.

After the processes have executed line~\ref{BITDESA-09}
(end of the first global synchronization phase),
each of them knows its  mapping function $\map_i()$,
but no process knows that all the other processes know their
own mapping function.
This motivates  the second use of the mutual exclusion algorithm,
which, as the left bit of any  register  $\SM[x].\BIT$ still contains
its initial value $0$,
ensures that when the last process (say $p_k$) that entered
the second critical section exits it, it knows that all the processes
have computed their mapping function, and
no process that executes the ``repeat'' loop of  line~\ref{BITDESA-16}
can exit it.

To identify the last process that entered the (second) critical section,
when a  process $p_i$ is inside the critical section it
increases the abstract register $\CT$ (line~\ref{BITDESA-11}),
and sets $last2_i$ to $\ttrue$ only if it discovers it is the last process
that accessed the critical section (line~\ref{BITDESA-12}),
More precisely, we have the following.
\begin{itemize}
\item
If $p_i$ is not the last process to increase $\CT$
(locally represented by $ct_i$), $last2_i$ is equal to $\ffalse$,
and consequently $p_i$ waits until it sees at least one
register whose bit $\SM_i[x].\BIT$ is equal to $1$ (line~\ref{BITDESA-16}).
When this occurs $p_i$ learns that the second phase is terminated
(hence it knows that all the processes have computed their mapping function),
and it can proceed to execute an upper layer non-anonymous register algorithm.
\item
Differently, if $p_i$ is  the last process to increase $\CT$,
it  changes to $1$ the left bit of all the registers
(line~\ref{BITDESA-15}), which unblocks all the other processes.
As the bits  $\SM_i[x].\BIT$ are never reset to $0$,
eventually all the processes know that each of them knows
its mapping function.
\end{itemize}
As they follow the same synchronization pattern,
the proof of the second part of Algorithm~\ref{algo:n-process-desa-final-v}
(lines~\ref{BITDESA-10}-\ref{BITDESA-17}) is the same as the one of
its first global synchronization phase
(lines~\ref{BITDESA-01}-\ref{BITDESA-09}), which is the same as the
one of Algorithm~\ref{algo:n-process-desanonymization}.

\section{Conclusion}
\label{sec:conclusion}
\vspace{-0.3cm}
In addition to introducing the memory desanonymization problem, this
paper has shown that, in an $n$-process system where $n\geq 2$ and
process identities can only be compared with equality, a shared memory
made up of $m$ anonymous read/write registers and a shared memory made
up of $m$ non-anonymous read/write registers have the same
computability power for the values of $m$ satisfying the
necessary condition for deadlock-free anonymous mutex
algorithms from~\cite{T17}, namely $m$ must belong to the set $M(n)=
\{~m~|~ \mbox{ such that }\forall \ell:~ 1 < \ell \leq n$:
$\ggcd(\ell,m)=1\} \setminus \{1\}$. Let us observe that, as it
includes an infinite sequence of prime numbers, $M(n)$ is infinite.
It follows that, once desanonymization  (in which all processes
participate) is obtained,
it becomes possible to use a symmetric starvation-free mutex algorithm,
thereby obtaining a symmetric starvation-free mutex
algorithm working on top of an anonymous memory
\footnote{Peterson's mutual exclusion
algorithm is such a  symmetric algorithm~\cite{P81}. As it  requires
$2n-1$ non-anonymous atomic registers, we need to have both
$m\in  M(n)$ and $m\geq 2n-1$.}.

We emphasize that the above construction (of running a starvation-free mutex algorithm on top of a desanonymization layer),
does not solve the original open problem from [20], regarding the existence of a memory-anonymous two-process starvation-free mutex algorithm. In the definition of the mutex problem participation is not required (a process may never leave its critical section), while our
implementation of the desanonymization layer, assumes that participation is required.

As stated in~\cite{T17},
the memory-anonymous communication model ``enables us to better understand the
intrinsic limits for coordinating the actions of asynchronous processes''.
It consequently enriches our knowledge of what can be (or cannot be) done
when an adversary replaced a  common addressing function, by
individual and independent addressing functions, one per process.

Among problems that remain open, there are  the design
of desanonymization algorithms (symmetric with equality only, or
symmetric with equality, greater than, and lower than)
not based on an underlying
memory anonymous mutex algorithm, and the statement of a
necessary and sufficient condition on the value of $m$ (size  of the
anonymous memory) for which desanonymization is possible
(for each type of symmetry).

\vspace{-0.1cm}
\section*{Acknowledgments}
\vspace{-0.1cm}
This work was partially supported by
  the French ANR project  DESCARTES (16-CE40-0023-03) devoted to layered
  and modular structures in distributed computing.
  ~\\~\\

\end{document}